\documentclass[a4paper]{jpconf}
\usepackage{graphicx}

\newcommand{\pbpb}{Pb--Pb}
\newcommand{\pT}{$p_{{\mathrm T}}$}
\newcommand{\jp}{J$/\psi$}
\newcommand{\raa}{$R_{\mathrm{AA}}$}
\newcommand{\sq}{$\sqrt{s_{\mathrm{NN}}}$}

\newcommand{\ee}{e$^+$e$^-$}
\newcommand{\mm}{$\mu^+\mu^-$}
\newcommand{\cc}{${\rm c}\bar{{\rm c}}$}
\newcommand{\bb}{${\rm b}\bar{{\rm b}}$}

\begin{document}
\title{\jp~Production~in~\pbpb~collisions~at~\sq=2.76~TeV}

\author{Ionut-Cristian Arsene for the ALICE Collaboration}

\address{Research Division and ExtreMe Matter Institute EMMI, GSI Helmholtzzentrum f\"{u}r 
Schwerionenforschung, Darmstadt, Germany}

\ead{i.c.arsene@gsi.de}


\begin{abstract}
We report on the \jp~production in \pbpb~collisions at \sq=2.76~TeV recorded with ALICE during the
2010 and 2011 LHC runs. The \jp~candidates are reconstructed using the \ee~and \mm~decay channels
at mid- and forward rapidity, respectively. We show the \jp~ nuclear modification factor \raa~as a 
function of centrality, transverse momentum
and rapidity. Our results are compared to the lower energy measurements from PHENIX and to 
model calculations.
\end{abstract}

\section{Introduction}
The Quark Gluon Plasma (QGP) is a state of deconfined nuclear matter formed in relativistic
heavy ion collisions. One of the promising proposed probes to study such a state is the suppression
of heavy quarkonia (bound states of \cc~and \bb~ pairs) via Debye screening \cite{matsui86}. 
Due to their large mass, the pair creation of heavy quarks occurs in
the initial partonic stage of the collision, thus enabling them to experience the entire
history of the collision. 
Without any cold nuclear matter (CNM) effect, the total production cross-section of these pairs 
in nuclear collisions should correspond to the one in nucleon-nucleon collisions scaled with the number 
of binary nucleon-nucleon collisions ($N_{\rm coll}$). 
It is expected that the production probabilities 
of quarkonium states are affected by the hot and dense QGP which makes them a sensitive tool for its
properties.
At the lower collision energies from SPS and RHIC, a strong \jp~suppression is observed. However,
the interpretation of these results is still under debate due to insufficient understanding of the
CNM effects which break the $N_{\rm coll}$ scaling. 
It is expected that the measurements in Pb--Pb collisions at LHC energies will shed more light on this subject.
Shadowing, the modification of the free nucleon gluon distribution functions, is the dominant contribution
to the CNM effects at LHC energies and will be measured in p-Pb collisions.

In these proceedings we report on measurements of the inclusive \jp~production as a function of
centrality, transverse momentum (\pT) and rapidity ($y$) in \pbpb~collisions at \sq=2.76~TeV
recorded using the ALICE setup during the 2010 and 2011 runs. The results are presented using the 
nuclear modification factor (\raa) which is defined as
\begin{equation}
R_{\rm AA}=\frac{1}{N_{\rm coll}} \times \frac{\sigma_{{\rm J}/\psi}^{Pb-Pb}}{\sigma_{{\rm J}/\psi}^{pp}}
\label{eq:raa}
\end{equation}
where $\sigma_{{\rm J}/\psi}^{Pb-Pb}$ and $\sigma_{{\rm J}/\psi}^{pp}$ are the inclusive \jp~cross-sections
in \pbpb~and pp collisions, respectively. $N_{\rm coll}$ is obtained using Glauber model calculations.

\begin{figure}[th]
\begin{center}
\begin{tabular}[htb]{cc}
\includegraphics[width=0.45\textwidth]{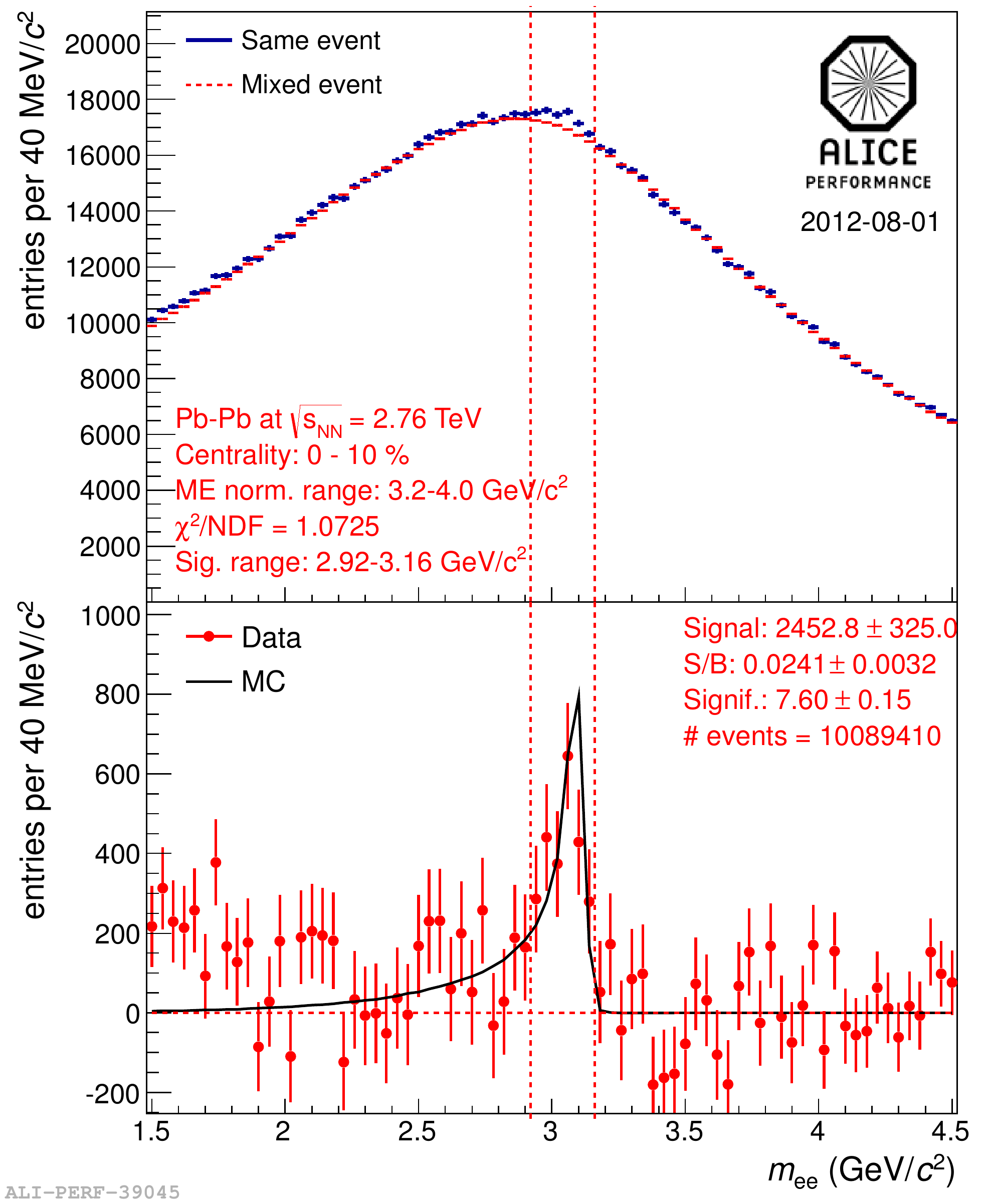}
&
\includegraphics[width=0.5\textwidth]{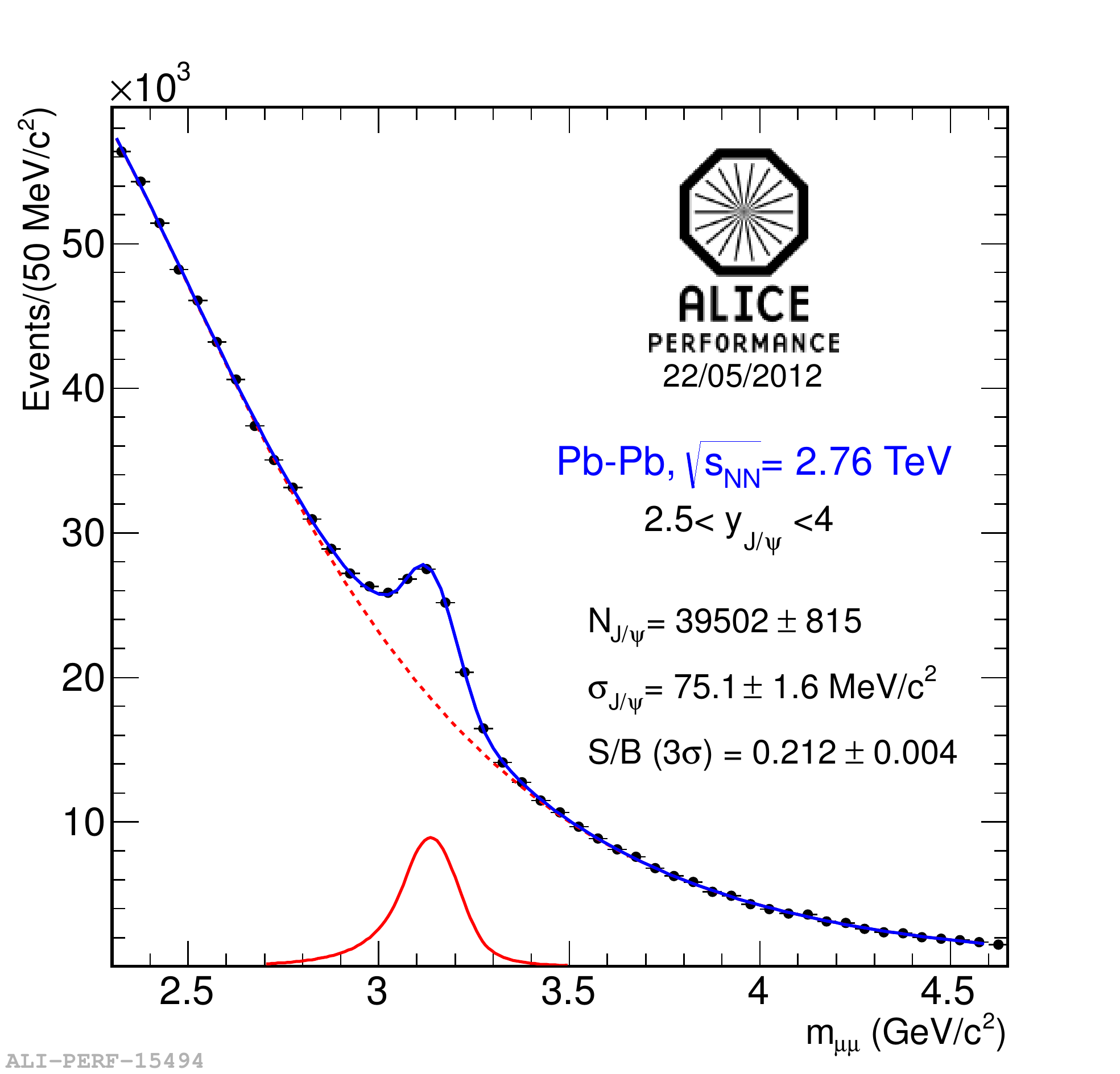} \\
\end{tabular}
\end{center}
\caption{Invariant mass distribution for \ee (left) in central (0-10\%) and \mm (right) pairs in minimum bias \pbpb~collisions.}
\label{fig:invmass}
\end{figure}

\section{Experimental setup and data analysis}

A detailed description of the ALICE experimental setup is provided in \cite{aliceJINST}.
It consists of several types of tracking and particle identification detectors covering 
the mid-rapidity region ($|\eta|<0.9$) and a forward muon spectrometer covering the range $2.5<\eta<4.0$.
Additional smaller detectors are used for triggering and event characterization.
At mid-rapidity, the \jp~are reconstructed in the di-electron decay channel. The electrons
are tracked using the Inner Tracking System (ITS) and the Time Projection Chamber (TPC). Electron identification
is done using the specific energy loss in the TPC gas.
At forward rapidity, the \jp~are measured in the di-muon decay channel, with the muons
being tracked, using five tracking stations, which are placed behind the hadron absorber.
A fast triggering system placed behind an additional iron wall ensures further background rejection
and increases the inspected luminosity.
The \jp~coverage at both mid-rapidity and forward rapidity extends down to 
zero transverse momentum. The present measurements do not allow for a separation between
direct and feed-down \jp~from the decays of higher mass charmonium states or from B mesons, so in the following we always refer 
to inclusive \jp.
Detailed descriptions of this analysis are given in \cite{jpsipp7,jpsipp276,jensHP2012,ionutQM2012,robertaQM2012}.

For the di-electron channel, the event sample used for this work consists of 30 million events,
the majority of them being centrality-triggered collisions. 
For the di-muon channel we used 17.7 million di-muon triggered events. The analyzed events correspond to an
integrated luminosity of 15$\mu b^{-1}$ in the di-electron channel and 70$\mu b^{-1}$ in the di-muon channel.

The \jp~signal is extracted from the invariant mass distribution constructed from opposite-sign (OS) pairs
of electron or muon candidates (see Fig.\ref{fig:invmass}). 
The dominant background contribution in the OS invariant mass distribution is coming from uncorrelated pairs.
For the di-electron channel, the background is constructed using the event mixing technique. 
Within the statistical uncertainties a good agreement is found between the background-subtracted distribution
and the signal shape as expected from our Monte-Carlo (MC) simulations (bottom-left panel of Fig.\ref{fig:invmass}). 
The contribution from the correlated
\ee~continuum surviving the mixed event background subtraction are incorporated in the systematic uncertainties 
for the signal extraction. Finally, the raw \jp~signal is obtained from bin counting in the range $2.92-3.16$~GeV/$c^{2}$.
In the di-muon channel, thanks to the dedicated trigger system and better signal to background ratio, the raw \jp~yields could be extracted
using two different methods. One method employs fitting of the \mm~distribution with a sum of a modified Crystal Ball (CB2)
function and a variable width gaussian for background parameterization (right panel of Fig.\ref{fig:invmass}). 
Alternatively, the mixed event background was subtracted
and the remaining distribution was fitted with the sum of a CB2 function for the signal and an exponential function
to parameterize the remaining correlated background. 

The raw \jp~signal was corrected for trigger efficiency, acceptance and reconstruction efficiency using 
simulated \pbpb~collisions employing the HIJING event generator enriched with primary \jp. 
The transport of the generated particles through the detector is done using GEANT3.
Furthermore, in the case of the di-muon decay channel, the efficiencies were calculated by embedding simulated 
\jp~into the raw data of real events. 
Details on the efficiencies and the systematic uncertainties can be found in 
\cite{jensHP2012,ionutQM2012,robertaQM2012}.

\begin{figure}[!t]
\begin{center}
\begin{tabular}[htb]{cc}
\includegraphics[width=0.5\textwidth]{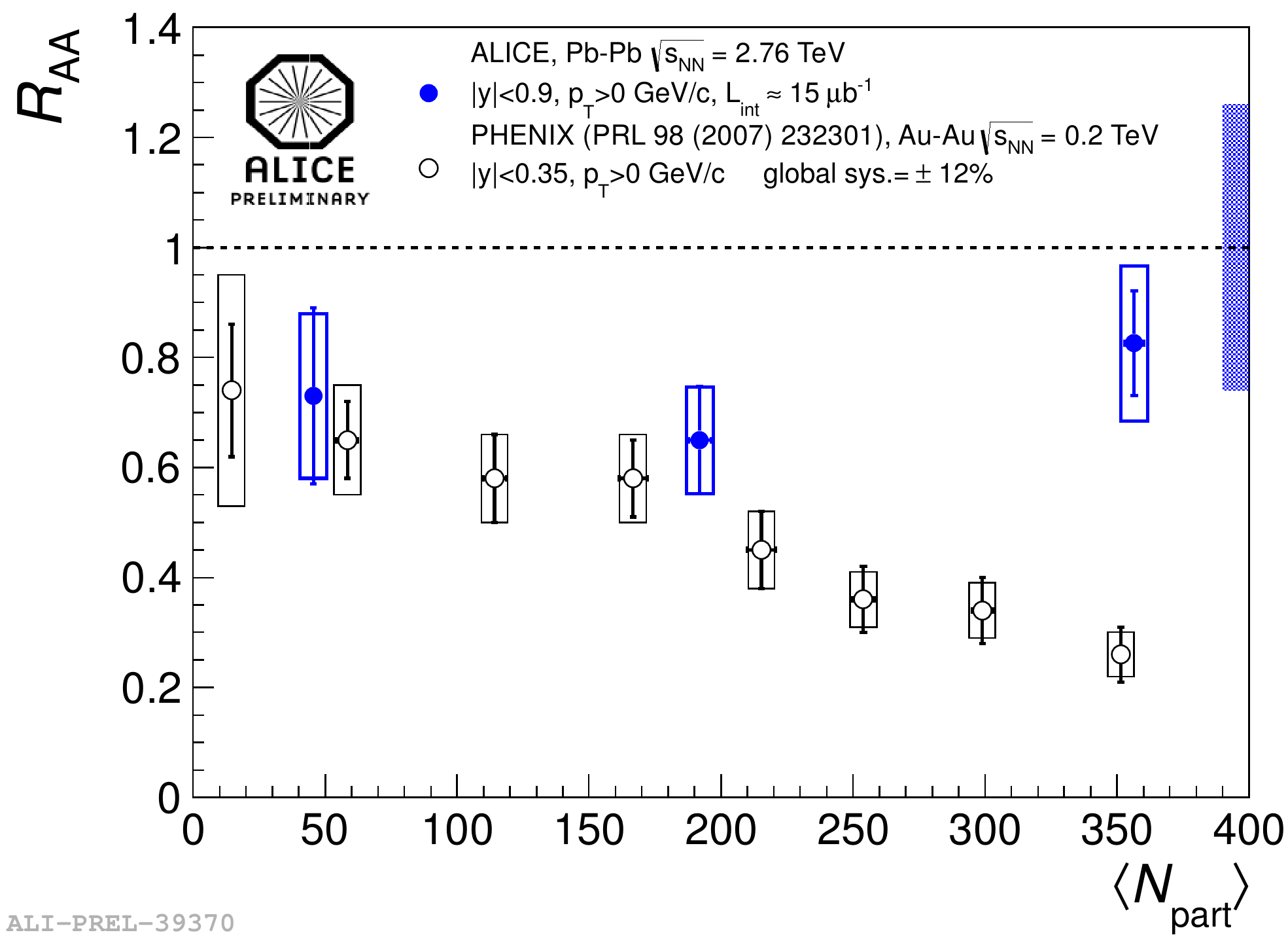}
&
\includegraphics[width=0.5\textwidth]{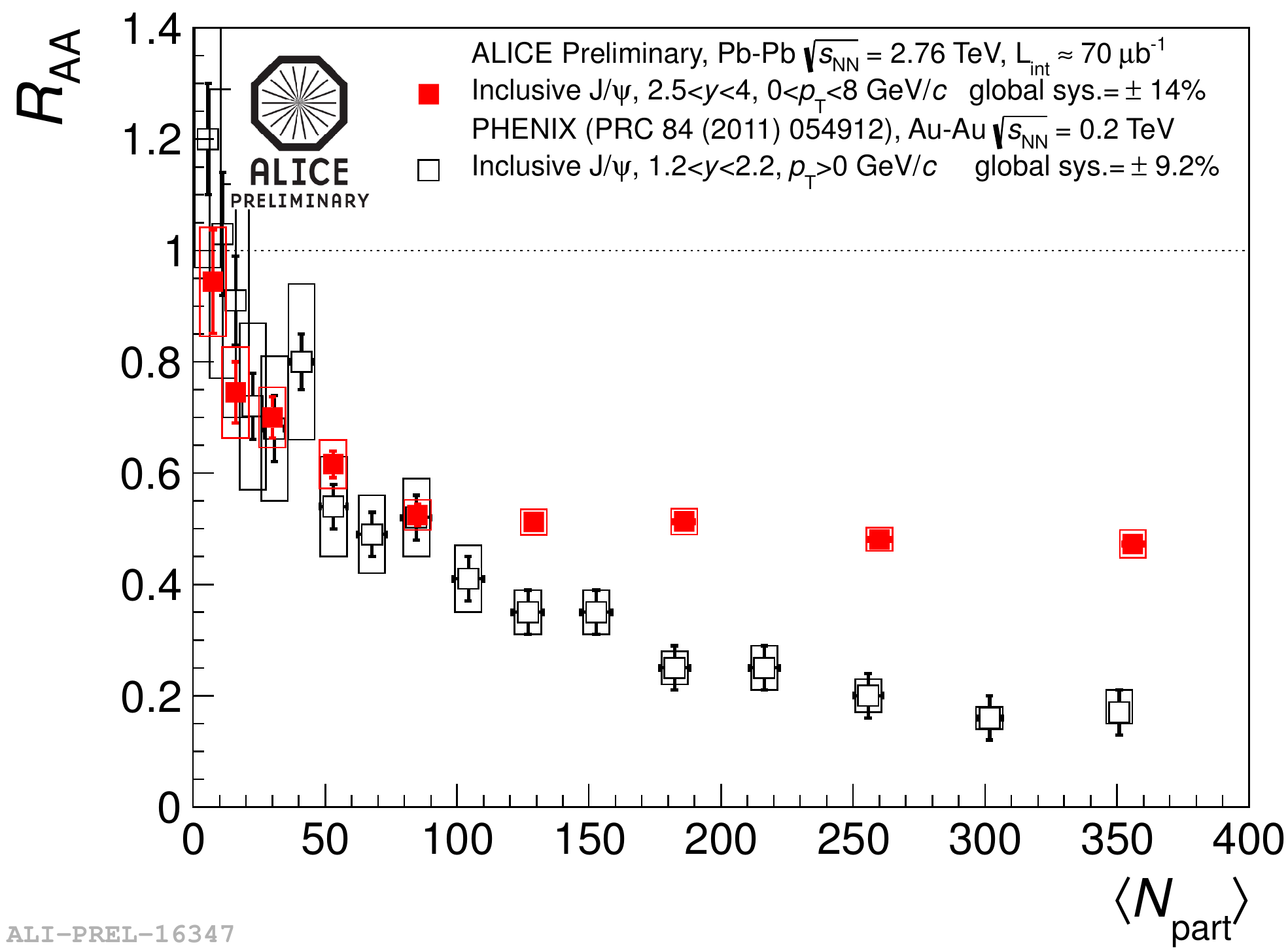} \\
\end{tabular}
\end{center}
\caption{\jp~nuclear modification factor \raa~as a function of the number of participant nucleons at mid-rapidity (left panel) 
  and forward rapidity (right panel) compared with the measurements from PHENIX in
  Au-Au collisions at \sq=200~GeV.}
\label{fig:raaCentPhenix}
\end{figure}

\section{Results and discussion}
The nuclear modification factor is obtained using the \jp~cross-section in pp
collisions (see Eq.\ref{eq:raa}) at $\sqrt{s}=$2.76~TeV measured by ALICE \cite{jpsipp276}. The systematic uncertainty on the \raa~
due to the statistical and systematic errors on the pp reference amounts to 26\% for the di-electron channel.
For the di-muon channel, the corresponding error, which is dominated by the systematic uncertainty, is 9\% for the 
integrated results and 6\% for the
\pT~or $y$ dependent results. 

Figure \ref{fig:raaCentPhenix} shows our results on the \raa~as a function of the number of participant nucleons ($N_{\rm part}$)
around mid-rapidity (left column) and at forward rapidity (right column). At mid-rapidity, the results show no $N_{\rm part}$
dependence within the experimental uncertainties and an \raa~value in the most central collisions of
$0.83\pm0.09({\rm stat.})\pm0.26({\rm syst.})$ has been obtained. At forward rapidity, 
the results indicate a clear \jp~suppression which is constant with centrality for $N_{\rm part}>$~50.
In Figure \ref{fig:raaCentPhenix} our results are compared with the ones from PHENIX measured in Au--Au collisions
at \sq=200~GeV \cite{phenix2007} at mid-rapidity (left) and forward rapidity (right). 
For the most central collisions, the \jp~\raa~is significantly higher at the LHC at both mid- and forward rapidity.
The \raa~measured by PHENIX shows a strong suppression in central collisions. Whether this observation
is compatible with \jp~suppression beyond the CNM effects, measured in d-Au collisions \cite{phenixDAu2011}, 
is still under debate. Phenomenological calculations from \cite{phenix2008} suggests
that the observed nuclear suppression in Au--Au collisions may be explained solely with the CNM effects, 
which include nuclear shadowing and the breakup of correlated \cc~pairs in cold nuclear matter.  
At the LHC, it is expected that the high charm quark density enhances the formation probability of quarkonia during or at the end 
of the QGP evolution \cite{pbm00, andronic07}. 
Our results for central collisions suggest that this effect could have a significant contribution. 

\begin{figure}[!t]
\begin{center}
\begin{tabular}[htb]{cc}
\includegraphics[width=0.5\textwidth]{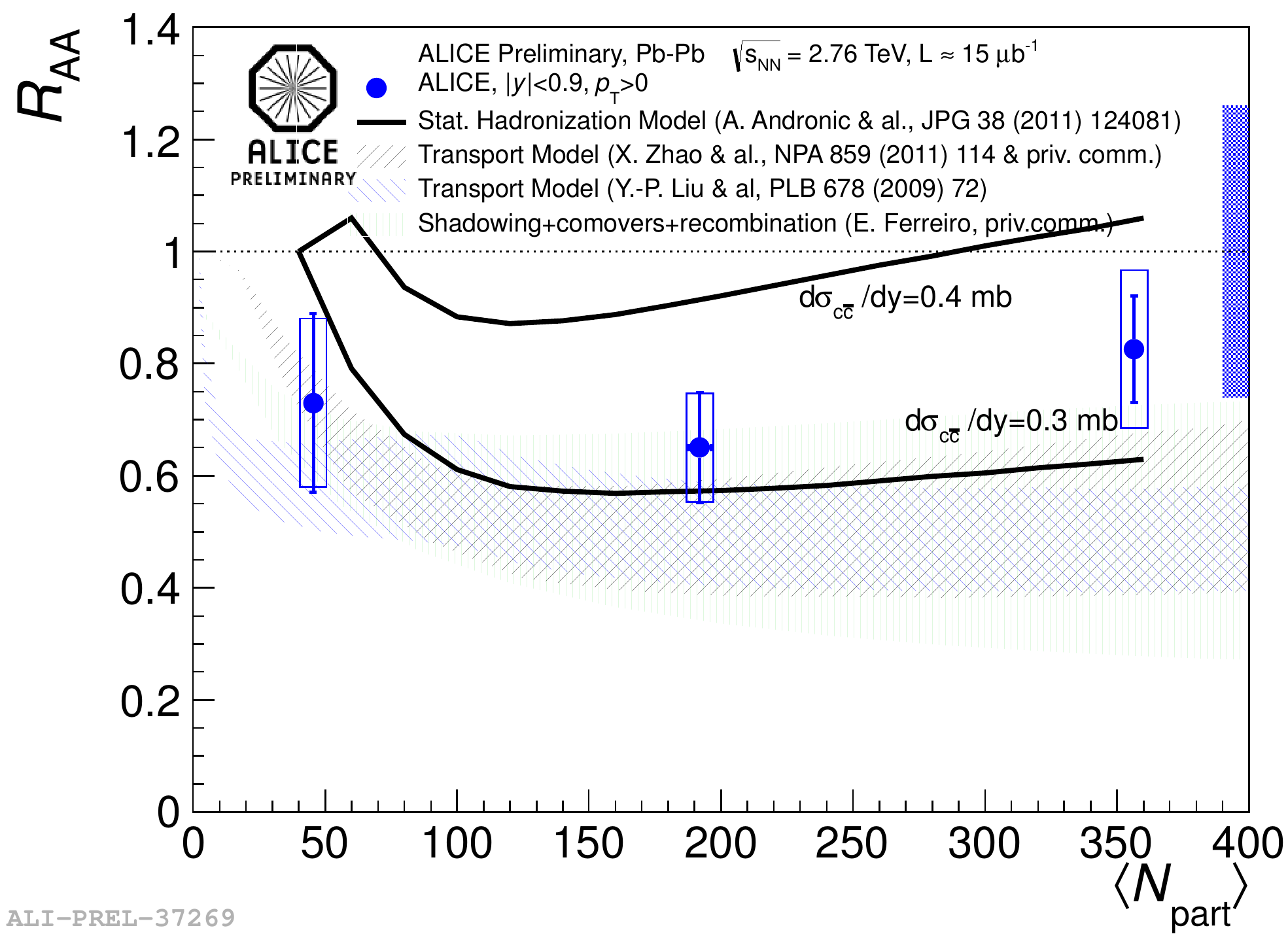}
&
\includegraphics[width=0.5\textwidth]{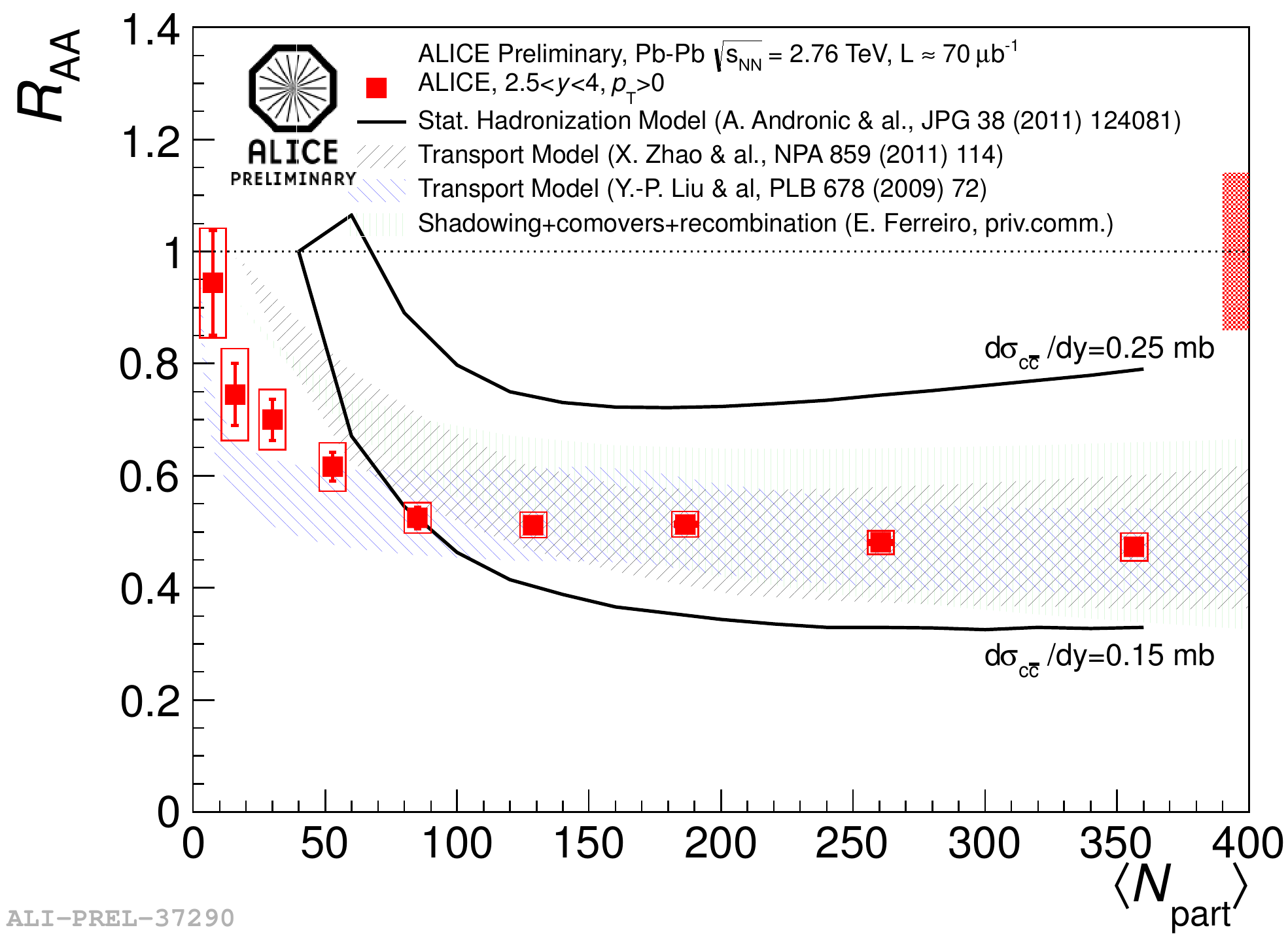} \\
\end{tabular}
\end{center}
\caption{\jp~nuclear modification factor \raa~as a function of the number of participant nucleons at mid-rapidity (left panel) 
and forward rapidity (right panel) compared with model calculations.}
\label{fig:raaCentModels}
\end{figure}

Figure \ref{fig:raaCentModels} shows a comparison of our results with model calculations which take this effect into account.
The hashed bands show the results from two transport models \cite{zhao2011,liu2009} and from the comover
interaction model \cite{ferreiro}. In these models
the fraction of \jp~resulting from (re-)combined \cc~pairs from the medium is at most 50\%, the rest being direct \jp~produced
during the hard partonic interactions. The solid lines show 
the prediction from the statistical hadronization 
model \cite{andronic2011}. This model assumes that no charmonium is formed in the QGP phase
and the charm quarks thermalize with the whole system. All the \jp~and other charmed hadrons 
are then formed at the chemical freeze-out and their yields can be calculated based on the total charm cross section 
and the thermal parameters. All of the model calculations have uncertainties due to the unknown inclusive 
$c\bar{c}$ production cross section and the CNM effects. The CNM effects will be measured in p-Pb collisions 
scheduled for beginning of 2013.

\begin{figure}[!tp]
\begin{center}
\begin{tabular}[htb]{cc}
\includegraphics[width=0.5\textwidth]{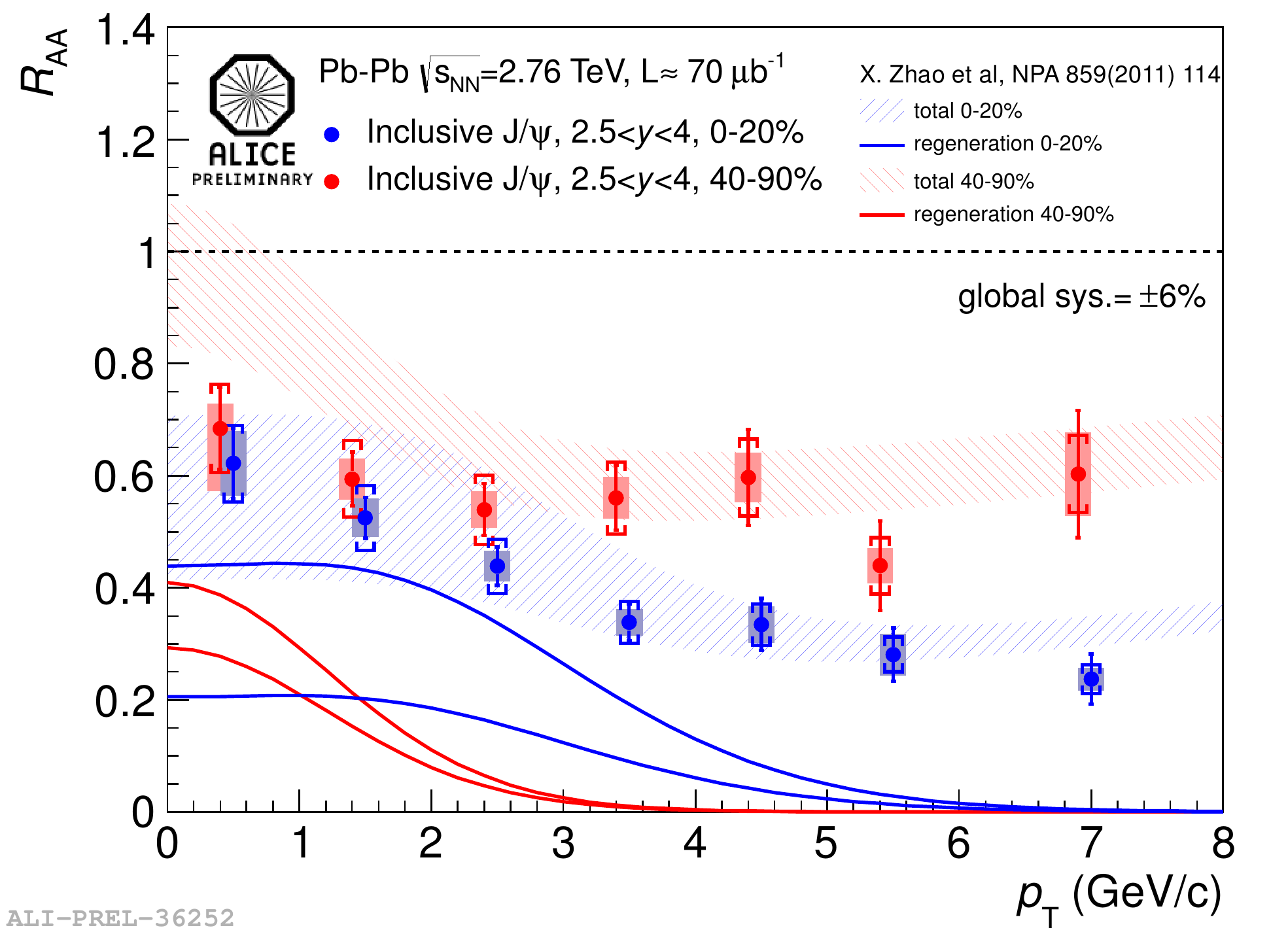}
&
\includegraphics[width=0.5\textwidth]{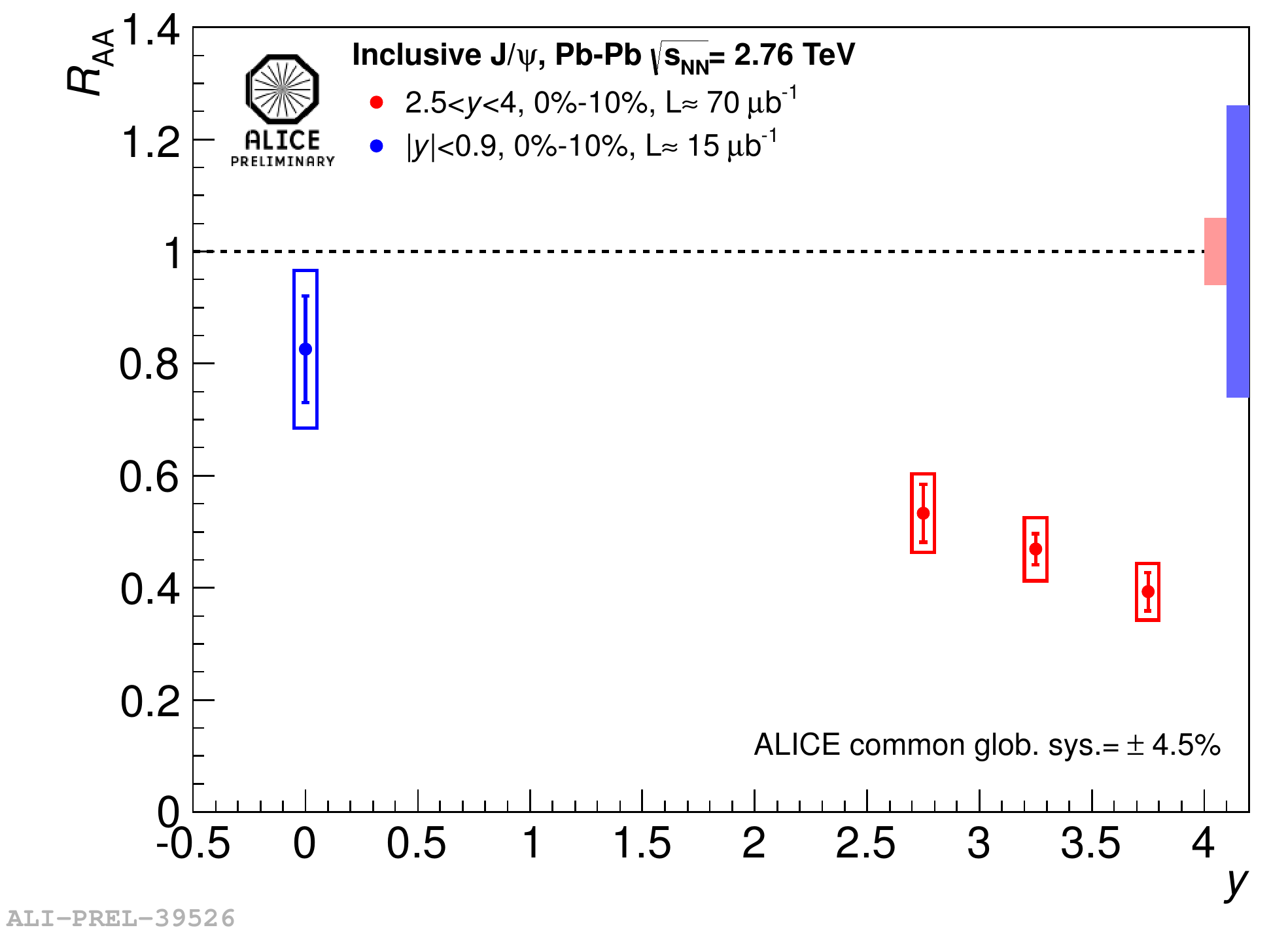} \\
\end{tabular}
\end{center}
\caption{\jp~ nuclear modification factor \raa~as a function of transverse momentum at forward rapidities ($2.5<y<4$) (left) and as a function of rapidity (right).}
\label{fig:raaPtY}
\end{figure}

Figure \ref{fig:raaPtY} shows the \pT~(left) and the rapidity (right) dependence of the \jp~\raa. The \pT~dependence
is shown for both central (0-20\%) and peripheral collisions (40-90\%) in comparison with calculations
from a transport model \cite{zhao2011}. The results for central collisions show 
that \raa~is $\approx$0.6 at low \pT~and decrease towards higher \pT. For the peripheral events,
we observe no or little \pT~dependence within the uncertainties. These results are in fair agreement with the model calculations from \cite{zhao2011}
with the exception of the peripheral collisions where the model seems to overestimate the data at low \pT.
The rapidity dependence of \raa~for the most central (0-10\%) collisions is shown in the right panel of Fig.\ref{fig:raaPtY}. 
We observe that the \raa~grows from forward to mid-rapidity. This behaviour was observed
also at RHIC energy in the most central collisions. Although there are more effects which need to be taken into account, 
these results might suggest that the (re)generation mechanism becomes important at LHC since 
the charm density is expected to grow from forward to mid-rapidity.


\section{Conclusions}
To summarize, we reported on the latest ALICE measurements of the inclusive \jp~nuclear modification factor as a function
of centrality, \pT~and $y$ in \pbpb~collisions at \sq=2.76~TeV. 
In the most central collisions our integrated \raa~values are roughly 3 times 
higher than the results obtained in central Au--Au collisions by PHENIX collaboration.
Model calculations which incorporate the (re)generation mechanism are in agreement with the data. Better knowledge 
of the total charm cross-section and the CNM effects will impose more stringent constraints to models.
A comparison of the \pT dependence of the \jp~\raa~in central collisions indicates that the (re)generation
mechanism is significant at low transverse momentum.
Finally, the \jp~\raa~in central collisions is decreasing with increasing rapidity as was also observed at RHIC.

\smallskip

\end{document}